\documentstyle[12pt,epsfig]{article}
\begin{document}
\def\figwidth{6.1in}
\def\beq{\begin{equation}}
\def\eeq{\end{equation}}
\def\bea{\begin{eqnarray}}
\def\eea{\end{eqnarray}}
\def\bq{\begin{quote}}
\def\eq{\end{quote}}
\def\ra{\rightarrow}
\def\lra{\leftrightarrow}
\def\la{\lambda}
\def\ups{\upsilon}
\def\eps{\epsilon}
\def\bq{\begin{quote}}
\def\eq{\end{quote}}
\def\ra{\rightarrow}
\def\lra{\leftrightarrow}
\def\un{\underline}
\def\ov{\overline}
\vspace*{-2.0cm}
\thispagestyle{empty}
\begin{flushright}
hep-ph/9810291\\
IOA-TH/98-10\\
CERN-TH-286/98
\end{flushright}

\vspace{1.5cm}
\begin{center}
{\large
{\bf 
\mbox{\boldmath{ $U(1)$}}--textures and 
Lepton Flavor Violation
}}
\end{center}

\vspace*{0.4 cm}

\begin{center}
{\bf M.E. G\'omez$^{a}$, G.K. Leontaris$^{a}$,
 S. Lola$^{b}$ and J.D. Vergados$^{a}$}
\end{center}

\begin{center}
\begin{tabular}{l}
$^{a}$
{\small Theoretical Physics Division, Ioannina University, 
GR-451 10 Ioannina,
Greece}\\
$^{b}$
{\small CERN Theory Division, CH-1211, Switzerland}
\end{tabular}
\end{center}

\vspace*{0.8 cm}

\begin{center}
{\bf Abstract}
\end{center}

\vspace{.5cm}\noindent
$U(1)$ family symmetries have led to successful predictions of
the fermion mass spectrum and the mixing angles of the hadronic
sector. In the context of the supersymmetric unified theories, 
they further imply a non-trivial mass structure for the
scalar partners, giving rise to new sources of flavor violation.
In the present work, 
lepton flavor non-conserving processes are examined in the context
of the minimal supersymmetric standard model
 augmented by a $U(1)$-family symmetry.
We calculate the
mixing effects on the $\mu\ra e\gamma$  and $\tau\ra 
\mu\gamma$ rare decays. All supersymmetric scalar masses involved
in the processes are determined at low energies using two
loop renormalization group analysis and threshold corrections.
{}Further, various novel effects are considered and found to
have important impact on the branching ratios. Thus, a rather
interesting result is that when the see-saw mechanism is applied
in the $12\times 12$ sneutrino mass matrix, the 
mixing effects of the Dirac matrix 
in the effective light sneutrino sector are canceled at first order.  
In this class of models and for the case that
 soft term mixing is already present at the
GUT scale, $\tau\ra \mu\gamma$ decays are mostly expected to arise at rates  
significantly smaller than the current experimental limits. On the other
hand, the $\mu \ra e \gamma$ rare decays  impose 
important bounds on the model parameters, particularly on the supersymmetric
scalar mass spectrum. In the absence of soft term mixing at high energies,
the predicted branching ratios for rare decays
are, as expected, well below the experimental bounds.

\vfill

\hspace*{0.5 cm}

\hrule\vspace{.3cm}  \noindent{\small
$^*$ Research partially supported by
the TMR contract ERBFMRX-CT96-0090.
}

\newpage

\section{Introduction}

In the last years there has been
a lot of interest in lepton flavor violation,
which can be a powerful tool in any
attempt to classify different extensions of the
Standard Model (SM) [1-4].
Indeed, there are various
ways to enlarge the known particle
spectrum in a manner that
lepton number violation is generated.
The new states may be
right-handed neutrinos,
additional vector-like heavy fermions,
supersymmetric particles, or even all the above
together as this is the usual case in supersymmetric 
grand unified theories (SUSY GUTS) and their string versions.

A well known result in the context of the non-supersymmetric standard
model is the conservation of lepton flavor in the case of zero
neutrino masses, while in the case of massive, non-degenerate
light 
 neutrinos, the amount of 
lepton flavor violation is proportional to the factor $\eta_{\nu} =\Delta
m^2_\nu/M^2_W$~\cite{P..}.  However, non-observation of double beta decay    
as well as other well known neutrino data, 
 imply severe bounds on the 
light neutrino mass-squared differences $\Delta m_{\nu}^2$,
leading to a high suppression
of the ratio $\eta_{\nu}$ and therefore to all flavor violating  processes
depending on it.
An analogous smaller suppression occurs if the process
is mediated by heavy neutrinos. When
supersymmetry enters in the game, the whole scene changes  completely.
Even in the absence of right handed neutrinos, flavor violations
could occur via the exchange of supersymmetric particles. A large
number of new parameters (sparticle masses, mixing angles, e.t.c.)
appear in the calculations, therefore enlarging  the
number of possible decays, while the   
predicted branching ratios are now comparable to  
the present experimental bounds.
Moreover, the soft breaking terms
may violate the individual
lepton numbers $L_e,L_\mu$ and $L_{\tau}$ by large
amounts. Therefore,
in the context of unification and low
energy phenomenology scenarios, 
flavor-violating  processes can provide useful
constraints on the parameter space
of a given model. 
Even in models with universal initial conditions for
the scalar masses, the renormalization group runs
of the slepton masses will give rise to lepton
flavor violation that can be significant.

Thus, there are various 
sources of Lepton Flavor Violation (LFV) in SUSY models
that may be tested in future searches
\cite{r}. Then, it is interesting to investigate them
in the framework
of flavor symmetries, which will give us a guideline
on the amount of flavor violation that we may expect
at the GUT scale. Indeed:
  
\begin{itemize}
 
\item
1) Unified supersymmetric theories with $U(1)$ family symmetries generate 
successfully the observed low energy hierarchy of the
   fermion mass spectrum and the quark mixing in terms of a minimum 
   number of arbitrary parameters at the unification scale. 
   In addition, a non-diagonal structure of the sparticle mass matrices
   comes out as a prediction of the theory. 
    Rare processes are  sensitive   
    to the scalar mass matrix structure and non-diagonality of the latter
    usually,
in a basis in which the fermions are
diagonal, leads to hard violation of flavor.
   
 \item
2) These flavor violating effects are enhanced in  particular when the 
    higgs vacuum expectation value (vev) ratio ($\tan\beta$) is large. 
     In fact, many models based
    on a single unified gauge group (like $SO(10)$ theory) predict 
    equality of the top and bottom Yukawa couplings at the GUT scale,
    and therefore a large value of $\tan\beta$ is implied. As a result, 
    the $6\times 6$ structure of the slepton mass matrix enhances 
    further the lepton mixing effects.

  \item
 3) When right-handed neutrinos enter in the model, the theory faces 
    another  challenge. Firstly, Dirac mass matrices arise of the order
    of the up-quark masses. In practice the majority of unified theories
    imply a Dirac matrix at the unification scale equal to the up quark
    matrix. Charged leptons and neutrinos are no longer diagonal in the
    same basis and a leptonic mixing matrix, similar to the Cabbibo
    Kobayashi Maskawa matrix
    $V_{CKM}$  for the quarks, is unavoidable. Secondly, the Dirac 
    mass matrix itself has an even more intriguing role, since,
    due to renormalization  effects on the supersymmetric scalar sector,
    it modifies the slepton mass matrix at low energies.
    Moreover, it enters 
     in the construction of the $12\times 12$- sneutrino mass matrix
     which in principle would have the potential to give rise to 
     additional flavor-violating effects. Nevertheless, since only the
     effective light sneutrino mass matrix is relevant in the calculation,
     we will show in this work that the $m_D$ effects are canceled at
     first order, when the see-saw mechanism is applied. 

 \item 
4) The contributions of the  trilinear scalar mass parameter of the potential, 
   i.e,  $A$-term contributions, are also discussed in this work in some detail.
  Due to  the absence of direct experimental information about its values,
  sometimes its effects are not considered at all. However, imposing even a zero 
 value for initial condition at the GUT scale, its low energy value will not  
 be unmarked.  Renormalization group running effects will drive its value to 
 magnitudes comparable with those of scalar masses, whereas its contribution to
 the branching ratios is of particular importance. 

\end{itemize}

In this work, we calculate the branching ratios for lepton flavor
violating decays in models with abelian flavor symmetries which
emulate the situation in many string constructions. We concentrate 
in particular in the class of  models for fermion
masses with one family symmetry which is the simplest possibility and have been 
firstly proposed in \cite{IR}.  This calculation 
offers an important  test with regard to the 
viability of these particular models,
but also gives further insight on the general
 predictions of family symmetries.
The motivation for their introduction is
 to explain the observed fermion mass  hierarchy  
yet they have further consequences as is the case of flavor violations.
The  lepton sector is ideal for such checks
(in the quark sector, large
uncertainties may enter in the calculations,
due to poor knowledge of hadronic matrix elements).

The analysis  follows the
lines of \cite{LT}, 
where some first
estimates for flavor-violation
in this class of models
have been presented.
This paper is organized as follows:

In section two we briefly analyze the basic features of a generalized class
of SUSY models with one $U(1)$-family symmetry. We give the forms of fermion
and scalar mass matrices and discuss their role in flavor violating parameters.

In section three, we derive the $12\times 12$ sneutrino mass matrix and show 
that the effective $3\times3$ light sneutrino matrix entering the flavor 
violating decays does not involve Dirac neutrino mass contributions.

In sections four and five we give the loop calculations for the various 
amplitudes and analyze the procedure for our numerical investigations.

In section six we present the results treating separately the cases with
and without scalar mass mixings at $M_{GUT}$. Finally, in section 7 we present
our conclusions.

\section{Mass matrices}

As has been stressed in the introduction, one of the main advantages
of the $U(1$)-family symmetries ($U(1)_f$) is the determination of the hierarchy
of the mass spectrum and mixing angles and the prediction of many other
parameters of the minimal supersymmetric standard model 
 (including the scalar matrices) with
only a minimal set of arbitrary parameters. In the simplest case of only
one $U(1)_f$ symmetry, the fermion mass hierarchy is successfully obtained
using an additional singlet field $\phi$ which develops a vev one order
of magnitude below the string scale $M_U$. 
Throughout our calculations, we will further assume the existence
of a  grand unified symmetry at a scale $M_{GUT}$
 without specifying the gauge group. (We have in mind models with
intermediate gauge symmetry groups\cite{alt} where the gauge couplings
run together from $M_U$ down to $M_{GUT}$.)  Below
the unification scale, only the minimal supersymmetric spectrum is assumed,
therefore the unification point will be taken at $\sim 10^{16}$ GeV.
In this scheme, the low energy parameters involved in our subsequent
calculations depend on $M_{GUT}$, the common value of the gauge 
coupling $g_{GUT}$ and the ratio of the singlet vev over the string
scale, $\epsilon\sim\langle\phi\rangle/M_U$.
 When the $U(1)_f$ symmetry is exact, only the third generation
has a Yukawa term in the superpotential, whilst all mixing angles are     
zero and lighter generations remain massless and uncoupled. When $\phi$
acquires a vev, the $U(1)_f$-symmetry is broken and mass terms fill in the rest
of the  mass matrix entries with Yukawa terms suppressed by powers of 
the ratio $\phi/M_U$.  In simple supergravity models, a
 similar situation occurs also in the 
scalar sector; at tree level, however, there appear three instead of
one, diagonal mass terms, one for each generation. Mixing (off-diagonal)
terms in the scalar sector shows up when higher NR-terms are included
in the K\"ahler potential. Thus, $U(1)_f$-symmetries imply also a non-trivial 
structure for the corresponding scalar mass matrices. This additional
structure may be  responsible for new hard flavor violations. 
The existing experimental bounds on flavor-violating processes
are therefore going to give us an indication as to
which structures are viable for the scalar mass matrices.

Which are the elements that determine the form
of the scalar mass matrices? Clearly, the charges of
the various fields under the flavor symmetry will play a dominant
role. Moreover, as we discussed in the
previous paragraph, the mass and mixing hierarchies
depend on expansion parameters that are
generated when singlet fields acquire vev's.
What these vev's can be, depends on the 
flat directions of a given theory.
Once the flat directions 
and the $U(1)_f$ charges of a particular
model have been fixed, the scalar mass matrix structure may be
easily computed through the K\"ahler function
${\cal G}={\cal K}+\log|{\cal W}|^2$
where ${\cal W}$ is the superpotential and ${\cal K}$ has the general form
\begin{equation}
{\cal K}=-\log(S+ S^*)-\sum h_n\log(T_n+ T_n^*)+
              Z_{ij^*}(T_n, T_n^*)Q_i Q_j^*+\cdots
\end{equation}
with $Q_i$ being the matter fields, $S$ the dilaton,
whereas $T_n$ are the other moduli fields.
The scalar mass matrices are determined by $Z_{ij^*}$ and ${\cal W}$. 
The form of the $Z_{ij^*}$ function is dictated by the modular symmetries
and depends on the moduli and the modular weights of the fields. Thus,
at the tree level, the diagonal terms are the only non-zero entries in
the scalar mass matrices. Higher order terms allowed by the symmetries
of the specific model fill in the non-diagonal entries.  

The lepton 
Yukawa interactions which appear in the superpotential 
in the presence of the right handed neutrino are
\beq
{\cal W}_{lep}= {e^c}^T\la_e\ell H_1 + N^c \la_D\ell H_2
 + \lambda_N \chi N^c N^c.
\eeq
Here $\ell$ is the left lepton doublet, $e^c$ is the right 
singlet charged lepton, $N^c$ is the  right-handed (RH) neutrino
and $\lambda_{e,D,N}$ represent  Yukawa
coupling matrices in flavor space.
Also, $H_1$ and $H_2$ are higgs doublets and $\chi$ stands for an effective
singlet which may acquire a vev at a large scale. 

In addition, soft supersymmetry
breaking terms generate  mass matrices for the charged slepton fields,
denoted by $\tilde m_\ell$, $\tilde m_{e_R}$. 
Denoting the various fields collectively with $z$, in  supergravity 
 the scalar potential is given by
\beq
V = e^{G(z)}\left(G_I G^{-1}_{I\bar{J}}G_{\bar{J}} -3\right)
+ \mid D\mid^2
\label{pt}
\eeq
where $\mid D\mid^2$ represents the contribution of the  $D$-terms in the
potential.
Also, with $G_I$, we denote the derivatives of $G$ with respect to the
fields $z_I$, i.e.,
\bea
G_I & \equiv
\frac 1{\cal W}{\cal D}_I{\cal W}
\eea
where ${\cal D}_I{\cal W}=\partial_I{\cal W}+{\cal W}\partial_I{\cal K} $
is the K\"ahler derivative.
Writing explicitly the various fields, in the low energy limit 
one encounters the following scalar mass terms  in $V$
\begin{eqnarray}
V &=& m_1^2  H_1^{*} H_1 + m_2^2  H_2^{*}  H_2 +
m_{\tilde q}^2 \tilde q^{*} \tilde q +
 m_{\tilde u^{c}}^2 {\tilde u^{c*}} {\tilde u^{c}}
  + m_{\tilde d^{c}}^2 {\tilde d^{c*}} {\tilde d^{c}} 
\nonumber \\ & + &
m_{\tilde \ell}^2 \tilde\ell^{*} \tilde\ell
 + m_{\tilde e^c}^2 \tilde e^{c*}\tilde  e^c
+ m_{\tilde N^c}^2 \tilde N^{c*} \tilde N^c
\nonumber\\
 &+& \{ \epsilon_{ab} (m_3^2  \tilde H_1^a \tilde H_2^{b} +
A_u \lambda_u \tilde q^a \tilde u^c \tilde H_2^{b}
 + A_d \lambda_d \tilde q^a \tilde d^c \tilde H_1^b  
\nonumber \\ & + &
A_l \lambda_e \tilde \ell^a \tilde e^c \tilde H_1^b +
 A_\nu \lambda_D \tilde \ell^a \tilde N^c \tilde H_2^{b}) 
+A_N \lambda_N \chi \tilde N^c \tilde N^c \})
+{\mathrm{ h.c.}} +
{\mathrm{ q.t.}} + \cdots \; ,
\end{eqnarray}
In the above equation
$a$ and $b$ are $SU(2)$ indices.
The dots stand for scalar 
trilinear {\it F}-terms  and q.t.
denotes quartic terms in the scalar fields.  Also terms 
proportional to $B$ parameter are not shown explicitly here.
The Higgs mass terms contain two contributions:
the first arising
from the superpotential and the second from the
soft supersymmetry breaking parameters: $m_i^2 = \mu^2 +
m_{H_i}^2$, with $i=1,2$.


We present here the relevant mass matrices of a  model whose
successful fermion mass hierarchy is predicted by $U(1)_f$ symmetries.
We work in the low energy effective model based on
the  $SU(3)\times SU(2)\times U(1)$
gauge group with an
additional $U(1)_f$ symmetry
\cite{IR}. 
After the implementation of this symmetry, the fermion matrix
for charged leptons in this model is given by
\begin{equation}
\label{eq:lmass}
{m_\ell}\approx \left (
\begin{array}{ccc}
\tilde{\epsilon}^{2|a +b|} &
        \tilde{\epsilon}^{|a|} &
              \tilde{\epsilon}^{|a +b|}
\\
\tilde{\epsilon}^{|a|} &
        \tilde{\epsilon}^{2|b|} &
             \tilde{\epsilon}^{|b|}
\\
\tilde{\epsilon}^{|a +b|} &
       \tilde{\epsilon}^{|b|} &1
\end{array}
\right){m_{\tau}}
\label{ml0}
\end{equation}
where the parameter $\tilde\epsilon$ is some power of the singlet vev scaled
by the unification mass, while $a,b$ are certain combinations of the
lepton and quark $U(1)_f$-charges. Order one parameters $c_{ij}$
in front of the
various entries (not calculable in this simple model) are assumed, to
reproduce the fermion mass relations after renormalization
group running. These parameters $c_{ij}$
are usually left unspecified, here however
their exact values are necessary for a reliable calculation of the 
lepton violating processes.  

A phenomenologically viable choice for the charges is to take
$a=3$ and $b=1$. A successful lepton mass hierarchy in this case is obtained
for the choice  $\tilde\eps = 0.23$. In this case, 
a possible choice of the coefficients $c_{ij}$ is given by
$c_{12} = c_{21} = 0.4, c_{22} = 2.2$, with the 
rest of the coefficients being unity.
We point out that the choice of the parameter $b$ is
not completely determined by the lepton mass texture. In fact there
is a second possibility, with $b=1/2$ where the fermion mass hierarchy 
is also consistent  with the low energy data.
On the contrary, the  choice of $b$, as well as the choice
of coefficients,  will have a significant impact on  the 
magnitude of the rare processes. Thus, flavor violation 
is  a powerful `tool' and an invaluable criterion of the viability  
of a certain choice. A detailed  discussion on this question
is one of the main points of the present analysis and 
will be given after the calculations on the branching ratios
will be presented in the subsequent sections.

The Dirac mass matrix in the above model has a similar structure. 
Due to the simple $U(1)$ structure of the theory, the powers
appearing in its entries are the same as the lepton mass matrix;
however, the expansion parameter is in general different\cite{DLLRS}.
Thus, its form is given by
\begin{equation}
\label{eq:dmass}
{m_{\nu_D}}\approx \left (
\begin{array}{ccc}
{\epsilon}^{2|a + b|} &
        {\epsilon}^{|a|} &
              {\epsilon}^{|a + b|}
\\
{\epsilon}^{|a|} &
        {\epsilon}^{2|b|}&
             {\epsilon}^{ |b| }
\\
{\epsilon}^{|a + b|} &
       {\epsilon}^{|b|}&1
\end{array}
\right){m_{top}}
\label{ml0b}
\end{equation}
The choice of charges $a=3, b=1$
 allows to identify the Dirac mass matrix with the up-quark mass
matrix.  Again, order one coefficients have to be introduced
in the quark mass matrix in order to obtain consistency with 
the experimentally determined masses. We denote here the corresponding
coefficients multiplying the entries of (\ref{ml0b}) with $d_{ij}$.
A choice of coefficients leading to correct up-quark masses
is obtained for  $d_{12}=
d_{21}=.5, d_{32}=d_{23}=1.5$, with the
rest of the coefficients being unity. In the case of
the up-quark matrix a  second 
expansion parameter\cite{IR} is introduced with the value
  $\epsilon = .053$.

The RH-Majorana mass matrix  is constructed from terms of
the form $\chi N^c N^c$ where $\chi$ is an effective
singlet. In  GUT models, $\chi$ is a combination of
scalar (Higgs) fields.
 Obviously, the structure of $M_N$ depends on the origin of 
the singlet $\chi$ as well as its charge\cite{DLLRS,NMIX2}.
 Thus, in a class of models this
singlet may arise from the combination $\tilde{N}^c
\tilde{N}^c$ where $\tilde{N}^c$ is the scalar component
of the RH-antineutrino supermultiplet. There are therefore
various structures of the Majorana matrix, depending on
the specific choice of the $\chi$-charge. 
However, as we are going to show analytically,
due to cancellations, the 
results are not sensitive to the
structure of $M_N$
\footnote{This, we checked numerically,
by picking particular forms 
of $M_N$, which also fit the light neutrino data.
For example, a zero singlet charge 
leads to a $M_N$ form similar to 
that of the Dirac mass matrix. 
 A singlet charge $Q= -1$
leads to an interesting form with large mixing in the 
2-3 generations and a two-fold degeneracy,
suggesting  a solution
of the atmospheric neutrino puzzle through the $\nu_{\mu}
\nu_{\tau}$ oscillations \cite{v12}. }.

The scalar mass matrices of this model are built using
the potential mentioned in the beginning of this section.
{}In particular, for the sleptons we obtain at the GUT scale
\beq
\label{eq:soft}
\tilde m^2_{\ell,e_R} \approx
\left (
\begin{array}{ccc}

{1} & \tilde\eps^{\mid a + 2 b\mid  }
&\tilde\eps^{\mid a + b\mid} \\
\tilde\eps^{\mid a + 2 b \mid } & {1} &
\tilde\eps^{\mid b \mid}\\
\tilde\eps^{\mid a + b\mid } & \tilde\eps^{\mid b\mid} & 1
\end{array}
\right)m_{3/2}^2
\eeq

At this point, we have determined all the necessary ingredients in order
to build the basic quantity which determines the flavor violations 
in SUSY theories.
 This is the  6$\times$6 slepton mass matrix which takes the form
\beq
\left(
\begin{array}{cc}
\tilde{m}^2_{\ell} & 
                       ((A_\ell+\mu\tan\beta)m_{\ell})^\dagger\\
(A_\ell+\mu\tan\beta)m_{\ell}  
                        & \tilde{m}^2_{e_R}
\end{array}
\right).
\label{6b6}
\eeq
$\tilde{m}^2_{\ell,e_r}$ are scale dependent, their value 
being given
by the renormalization group equations
Their initial conditions  at $M_{GUT}$  are
defined by the matrices (\ref{eq:soft}) and have a significant
 effect in flavor violations.  In addition, as it has been
extensively discussed in the literature, due to RGE running
slepton mass matrices receive corrections proportional to the
Dirac mass matrix. Charged lepton and Dirac mass matrices are
not simultaneously diagonalized; thus in the basis where $m_{\ell}$
is diagonal, the $3\times 3$ slepton mass matrix acquires a
non-diagonal contribution of the form
\bea
\delta\tilde{m}_{\ell}^2\propto \frac 1{16\pi^2} (3 + a^2)
\ln\frac{M_{GUT}}{M_N}\lambda_D^{\dagger} \lambda_D m_{3/2}^2
\eea
where $\lambda_D$ is the Dirac Yukawa coupling and 
the proportionality factor depends on the scalar mass
parameters squared while the parameter $a$ is 
related to the trilinear mass parameter  $A_l= a m_{3/2}$.
Thus, even if one starts with a diagonal slepton mass matrix,
there are important off-diagonal contributions due to the
existence of $N^c$.  This effect, however, will prove less
important in theories where the scalar mass matrix textures are 
also determined by $U(1)$ symmetries.

\section{Sneutrino mass matrix}

The sneutrino mass matrix is also determined similarly. 
It is a $12\times 12$ structure given in terms of the $3\times 3$
Dirac,  Majorana  and slepton mass matrices.
It is generally expected that --as in the case of charged sleptons--
the Dirac term induces considerable mixing effects. We will
show here that this is {\it not} the case in the sneutrino mass matrix. 

 This $12\times 12 $ matrix 
is rather complicated and not easy to handle.
Vastly different scales are involved and numerical investigations
should be carried out with great care.
Its form is as follows:
\bea
\label{eq:tw}
\begin{array}{c| c c c c c}
&  \tilde{\nu}  &  \tilde{\nu}^{\ast}  &  \tilde{N}^c&   {\tilde{N^c}}^\ast\\

\hline
 \tilde{\nu}^{\ast} & m_{\tilde{l}}^2+ m_D^{\ast} m_D^{T}  & 0 &
m_D^{\ast} M^T & \left({A^{\ast}_\nu}+ \mu cot\beta\right) m_D^{\ast}\\
& & & & \\
 \tilde{\nu} & 0 & m_{\tilde{l}}^2+ m_D m_D^{+} &
\left({A_{\nu}}+ \mu cot\beta\right) m_D & m_D M^+\\
& & & & \\
 {\tilde{N^c}}^{\ast} &
M^{\ast} m_D^T & m_D^+ \left({A^{\ast}_\nu}+ \mu cot\beta\right) &
m_N^2+M^\ast M^T 
&{A_N}^\ast M^\ast\\
& & & +m_D^\ast m_D^+ & \\
& & & & \\
 \tilde{N^c} &
m_D^T \left({A_{\nu}}+ \mu cot\beta\right) & M m_D^{+} & A_N M &
m_N^2+M M^+ \\
& & & &  +m_D m_D^+ \\
\end{array}
\eea

One can construct an effective $6 \times 6$ matrix
for the light sector, by applying matrix
perturbation theory, similar to the see-saw mechanism.
The result up to second order is: 

\beq
(m_{\tilde{\nu}}^2)_{eff} =
\left(
\begin{array}{cc}
m_{\tilde{\ell}}^2 -                    
(A_\nu+\mu\cot\beta)(A_\nu - 2 A_N) \cdot
 &
 ( (2 A_\nu + A_N) + 2 \mu \cot\beta) \cdot \\
\hspace{0.9 cm}
(m_D M^{-2} m_D^\dagger) & 
\hspace{0.9 cm}
(m_D M^{-1} m_D^\dagger) \\
  & \\
 ( (2 A_\nu + A_N) + 2 \mu \cot\beta)\cdot &
m_{\tilde{\ell}}^2 -                    
(A_\nu+\mu\cot\beta)(A_\nu - 2 A_N)\cdot \\
\hspace{0.9 cm}
(m_D M^{-1} m_D^\dagger)^* &
\hspace{0.9 cm}
(m_D M^{-2} m_D^\dagger)
\end{array}
\right)
\eeq
The first and second order terms are obtained assuming all parameters
as real and the $A-$matrices proportional to the identity. 
Notice that the second order terms along the diagonal
can be neglected. The first order off-diagonal
terms must be retained, since they lead to
complete mixing of the pairwise degenerate states.
This, however, does not affect the flavor-violating 
branching ratio.

The simplicity of this result is rather astonishing. We note that
after the `see-saw' mechanism is applied, 
the Dirac neutrino mass matrix contribution in the
effective light snetrino mass sector is essensially negligible.
Moreover, there is an additional benefit, since the complication of
the initial $12\times 12$ mass matrix can now be avoided. 
A direct numerical calculation
of mass eigenstates and mixing angles
would be a hard task, due to the vastly
different scales. 

\section{Amplitudes for flavor violating processes}

{}Figure 1 shows the one-loop diagrams relevant to the $\mu\ra
e\gamma$ process. The corresponding $\tau\ra \mu\gamma$-decay is
represented by an analogous set of graphs. There are 
also box-diagrams contributing to this process; they
are however relatively suppressed.

The electromagnetic current operator between two lepton  states $l_i$
and $l_j$ is given in general by
\begin{eqnarray}
{\cal T}_\la &=& \langle l_i(p-q)|{\cal J}_\la|l_j(p)\rangle\nonumber\\
{  }&=&{\bar u_i}(p-q)
      \{ m_j i\sigma_{\la\beta}q^\beta 
               \left(A^L_MP_L+A^R_MP_R\right)
      \} u_j(p)
\label{general}
\end{eqnarray}
where $q$ is the photon momentum. The $A_M$'s  have
contributions from neutralino-charged slepton ($n$) and
chargino-sneutrino ($c$) exchange
\begin{equation}
A_M^{L,R}=A_{M(n)}^{L,R}+A_{M(c)}^{L,R}
\label{ampl}
\end{equation}
The amplitude of the process is then proportional to ${\cal T}_\la
\epsilon^\la$ where $\epsilon^\la$ is the photon polarization vector.
An easy  way to determine the loop momentum integral contribution
to the $A_M$'s is to search, in the corresponding diagram, for terms
of the form $(p\cdot\epsilon)$ and make the replacement
$2(p\cdot\epsilon) \ra \imath \sigma_{\la\beta}q^{\la}\epsilon^{\beta}$. 
Defining the ratio
$x=M^2/m^2$, where $M$ is the chargino (neutralino)  mass and $m$ the
sneutrino (charged slepton) mass, the following functions appear in
the $A_M$ term
\begin{equation}
\begin{array}{lll}
A_{M(n)}:\quad&\frac{1}{6(1-x)^4}(1-6x+3x^2+2x^3-6x^2\log x) & 
({\rm L-L \;  amplitude})\\
        &\frac{1}{(1-x)^3}(1-x^2+2x\log x)\sqrt{x} & (
{\rm L-R \;  amplitude})\\
A_{M(c)}:\quad&\frac{1}{6(1-x)^4}(2+3x-6x^2+x^3+6x\log x)& 
({\rm L-L \;  amplitude})\\
        &\frac{1}{(1-x)^3}(-3+4x-x^2-2\log x)\sqrt{x} 
& ({\rm L-R \;  amplitude})
\end{array}
\end{equation}
where $m_{l_j}$ is the mass of the $l_j$ lepton.

Notice in the L-L amplitudes
the lack of terms proportional to the gaugino mass $M$ 
which cancel. 
The Branching Ratio ($BR$) of the decay $l_j\ra l_i+\gamma$ is given by
\[
BR(l_j\ra l_i\gamma)=\frac{48\pi^3\alpha}{G_F^2}
               \left((A_M^L)^2+(A_M^R)^2\right)
\]

\section{Inputs and Procedure}

The branching ratio formulae for 
the $\mu\ra e\gamma$ and $\tau\ra \mu \gamma$ decays involve the masses
of  most of the supersymmetric particles. It is important therefore
for any given set of GUT parameters to know precisely all masses and 
the other low energy parameters. In the present work, 
 this is obtained by numerical integration
of the renormalization group equations of the MSSM with right handed neutrinos.
The  renormalization-group
equations can be found in many papers (see for example\cite{rge}).

We evaluate the coupling constants, using renormalization-group
equations at two loops.
Threshold effects are also taken
into account, by decoupling every sparticle
at the scale of its running mass $Q=m_i(Q)$.
Below  the scale $m_t$, we use the SM  beta functions..
                                              
Our analysis uses as input  
values  the unified coupling constant $\alpha_G$, at
the GUT  scale,
the third generation Yukawa couplings $\lambda_{t}$, $\lambda_{b-\tau}$,
the common scalar diagonal scalar masses $m_{3/2}$,  the
gaugino mass $m_{1/2}$, the (effective) Higgs bilinear coupling $\mu$,
and the ratio of the Higgs vev's
described by $tan\beta$ and  the flavor-symmetric
soft-breaking parameter $A_0$.

Our integration procedure consists on iterative runs of the
renormalization-group equations from
$M_{GUT}$ to low energies and back, for every set of input parameters
$m_{1/2},m_{3/2}, A_0$ and $tan\beta$,
 until agreement with experimental
data is achieved.   
The values for $\alpha_G$ and $M_{GUT}$ are obtained
consistently with $\alpha_{em}, \alpha_3$
and $sin^2{\theta_W}$ at $m_Z$.
Supersymmetric corrections to $sin^2{\theta_W}$ are
also considered\cite{sin2} .

A similar procedure is followed
to obtain the GUT values for the third
generation Yukawa couplings.
In particular, the GUT value for $\lambda_t$ is adjusted
by requiring the top physical
mass to be $m_t=175 \pm 5 GEV$. Similarly, we
obtain the
value of the unified $\lambda_{b-\tau}$,
by requiring the
correct prediction for $m_\tau=1.778 GEV$.
In all the cases that we analyzed, values
for $m_b$ consistent with experiments are found once QED and QCD corrections
are taken into account \cite{larin}.

The value of the $\mu$ parameter    
(up to its sign) can be expressed in
terms of the other input parameters by means of symmetry breaking conditions.
To this end, we use
the semi-analytic formulae including one loop corrections,
as given in ref.\cite{leonta}.
Finally, the chargino and neutralino
masses are obtained by
diagonalization of the $4 \times 4 $ neutral and the $2\times2$ charged
matrices as described in \cite{Gun}.
The RG evolution of the eigenstates, starting from
universal initial conditions at the GUT scale,
is properly taken into account.

We then explore the values of
the $BR$ for all significant values of the input
parameters $m_{1/2}, m_{3/2}$, $ A_0$ and $tan\beta$.

If we consider common scalar masses and trilinear terms at the GUT scale,
leptons and sleptons will be diagonal in the same superfield basis. However,
due to the presence of (a) the non diagonal GUT terms
$\Delta$  at the GUT scale,
and  (b) the appearance of
$\lambda_D$ in the RG equations, the  lepton Yukawa
matrix  and the slepton mass matrix  can not be brought simultaneously
to a diagonal form at the scale of the heavy Majorana masses.
Therefore, lepton number
will be violated by the one loop diagrams of fig. 1.

We define the unitary matrices diagonalizing the Yukawa mass textures
$\lambda_D$ and $\lambda_e$, as follows
\begin{eqnarray}
\label{eq:diag}
\lambda_D^\delta&=&T_R^T \lambda_D T_L \\
\lambda_e^\delta&=&V_R^T \lambda_e V_L
\end{eqnarray}
Here, the index $\delta$ indicates
a diagonal form. Then, the mixing
matrix $K$ in the lepton sector, defined in   
analogy to $V_{CKM}$  is given by the product
\begin{equation}
K=T_L^\dagger V_L
\end{equation}
The charged slepton masses are obtained by numerical diagonalization of
the $6 \times 6 $ matrix
\begin{equation}
\label{eq:66}
\tilde{m}_e^2=\left(\begin{array}{cc} m_{LL}^2&m_{LR}^2\\
                                   m_{RL}^2&m_{RR}^2 \end{array}\right)
\end{equation}
where all entries are $3 \times 3$ matrices in the flavor space.
In the superfield basis where $\lambda_e$ is diagonal, it is convenient 
for later use to write the $3\times 3$ entries of (\ref{eq:66}) in the form:
\begin{eqnarray}
m_{LL}^2&=& (m_{\tilde{l}}^\delta)^2+ \delta m_N^2+\Delta_L+m_l^2 +M_Z^2       
(\frac{1}{2} -sin^2\theta_W) cos 2\beta\\
m_{RR}^2&=& (m_{\tilde{e_R}}^\delta)^2+\Delta_R+m_l^2
 -M_Z^2 sin^2\theta_W cos 2\beta \\
m_{RL}^2&=& (A_e^\delta +\delta A_e + \mu tan\beta) m_l\\
m_{LR}^2&=& m_{RL}^{2\dagger}
\end{eqnarray}
Each component  above has a different origin and gives an  independent contribution 
in the Branching Ratios.
We further wish to emphasize the following:
\begin{itemize}
\item $ (m_{\tilde{l}}^\delta)^2, (m_{\tilde{e_R}}^\delta)^2, A_e^\delta$ 
denote the scalar diagonal contribution of the corresponding matrices;
their  entries are obtained by numerical integration of the
RG equations as described before. We consider $m_{3/2}^2$ as the common
initial condition for the masses at the GUT scale, while 
the trilinear terms scale as $a m_{3/2}$.
Since in the RGEs
we consider only third generation Yukawa couplings and common
initial conditions at the GUT scale for the soft masses, our treatment is
equivalent to working in
superfield basis, such that:
(i)  $\lambda_D$ is diagonal
from the GUT scale to the intermediate scale and
(ii) $\lambda_e$ is diagonal from the intermediate scale to low energies.
The change of bases will produce a shift  in the diagonal elements of the soft
mass matrices
at the GUT and at the intermediate scale.
This effect is negligible
(less than one percent).

\item  $\delta m_N^2$ and $\delta A_{l}$ stand for  the
off-diagonal terms which appear due to
the fact that
$\lambda_D$ and $\lambda_e$ may not be diagonalized
simultaneously.
The intermediate scale 
that enters in the calculation
(which is the mass scale for the neutral Majorana
field $M_N$) is defined by demanding that neutrino masses $\approx 1 eV$
are generated via the
``see-saw'' mechanism. This sets the $M_N$ scale 
to be around the  value $10^{13}$ GeV.
Then, the following values are obtained:
\begin{eqnarray}
\delta m_N^2&=& K^\dagger \left[ m_{\tilde l}^2(m_N)\right]K|_{ non diagonal}
\label{dN}\\
\delta A_l&=& V_L A_l(m_N) V_L^\dagger |_{ non diagonal}
\label{dA}
\end{eqnarray}

\item
The following values for
$\Delta_L$ and  $\Delta_R$ are defined at the GUT scale:
\begin{eqnarray}
\Delta_L&=& V_L^\dagger  \Delta V_L\\
\Delta_R&=& V_R^\dagger  \Delta V_R
\end{eqnarray}
\end{itemize}      
The effective $3  \times 3$
sneutrino mass matrix squared has the same form as the $m_{LL}^2$
part of the $6 \times 6$ charged slepton one,
with the difference that now Dirac masses
are absent (in consistency with
what we have shown in the analysis of the
$12 \times 12$ sneutrino matrix). Thus,
\begin{equation}
\tilde{m}_{\nu}^2= (m_{\tilde{l}}^\delta)^2+ \delta m_N^2+\Delta_L + \frac{1}{2}
M_Z^2 cos 2\beta
\end{equation}

It is illustrative to write our results as an approximate function of the
input parameters. Below we give the numerical range for the sneutrino and
the $A$ parameter as these are defined in (\ref{dN}),(\ref{dA}):

\begin{eqnarray}
\delta m^2_N \approx
\left (
\begin{array}{ccc}
0 & (4.2-6.3) \times 10^{-5} 
& (2.3-3.3) \times 10^{-4} \\
(4.2-6.3) \times 10^{-5} 
& 0 & (0.7-1.1) \times 10^{-2} \\
(2.3-3.3) \times 10^{-4} &
(0.7-1.1) \times 10^{-2} & 0 
\end{array}
\right )
(3+a^2) m^2_{3/2}
\end{eqnarray}

\begin{eqnarray}
\delta A_{\ell} \approx
\left (
\begin{array}{ccc}
0 & (1.2-1.7) \times 10^{-4} &
(5.2-7.3) \times 10^{-4} \\
(1.2-1.7) \times 10^{-4} &
0 & (1.0-1.4) \times 10^{-2} \\
(5.2-7.3) \times 10^{-4} &
(1.0-1.4) \times 10^{-2} & 0 
\end{array}
\right ) A_0
\end{eqnarray}

In the last two equations, the ranges in parentheses
correspond to $\tan\beta$ values between 14 and 3
(larger contributions are obtained for
smaller values of $\tan\beta$).         

\section{Results}

We have seen that, when  $U(1)$-family
symmetries are taken into account
 non-diagonality in the mass matrices is
generic in  both the fermion and the 
scalar sector.
This may generate
unacceptably large flavor-violating effects.
It is possible that cyclic permutation symmetries between generations
and universal anomalous $U(1)$-factors may prevent mixing effects
in the supersymmetric mass matrices \cite{AP}.
In our results, we are considering separately two
distinct cases: 
{}First, we will consider the case where the scalar mass matrices
are protected from mixing effects by some kind of symmetry not
affecting the fermion mass sector. Second, we will allow mixing
effects in both sectors, and recalculate the branching ratios
and the new bounds obtained on the sparticle spectrum.

{\bf 1). Case without scalar mass mixing at the GUT scale, $\Delta =0$.}

We start with the process $\mu\ra e\gamma$,
in the absence of mixings at the GUT scale.
Let us denote by  $(n)$ the 
contributions from the neutralino-charged slepton
exchange,
and by $(c)$ the ones from the 
chargino-sneutrino.
Then,
\bea
A_M^{L,R}= A_{M(n)}^{L,R}+ A_{M(c)}^{L,R}.
\eea

The various amplitudes appear in Fig.2.
As we can see, the two contributions to $A_{M}^R$ (dashed lines)
are of 
the same order of magnitude and opposite signs, while
their magnitude decreases
with $\tan \beta$. Around a certain 
value of $m_{3/2}$, there is a partial cancellation
of both amplitudes, leading to a  decrease of the   
expected $BR(\mu\rightarrow e+\gamma)$, as we 
can see in Figs.\ref{fig:offm},
\ref{fig:offtb} and \ref{fig:offa}.

The contribution to $A_{M}^L$ comes almost exclusively from  $A_{M(n)}^{L}$,
since the chargino exchange contribution  $A_{M(c)}^L$
(Feynman diagram b in Figure 1) arises
due to Yukawa interactions and is about three orders of magnitude smaller
than the other contributing amplitudes. It is important to make this remark, 
since when there are no mixings in the right-handed slepton masses,
  $A_{M(n)}^L$ become relevant due to the presence of non-diagonal 
mixings from the trilinear terms $\delta A $ (otherwise,
these contributions are of the same
order as $A_{M(c)}^L$). The effect of $\delta A $ can be
seen clearly in Fig. \ref{fig:offa}. 
For the initial condition $A_0=0$, $\delta A $ is significantly suppressed,
and hence a dramatic  decrease of the $BR(\mu\rightarrow e+\gamma)$ is observed.
On the other hand,  the cases  with
$A_0\neq 0$ have a  remarkable difference with the previous one. 
All  curves now are smoother while there are no  particular $m_{3/2}$-values
where $BR$ exhibit large suppression.

  Although we start in our case with universal soft masses at the GUT scale,
there is an analogy with the situation of the $SU(5)$ model
discussed elsewhere \cite{hisano,bh2} in the following sense.
Assuming we are in a basis where Yukawa matrices are diagonal,
the renormalization of the universal soft mass terms from
$M_{GUT}$ down to the RH-neutrino mass scale $M_N$ will split
mass parameters of different generations. As a result, flavor
universality in the scalar sector at $M_N$ is lost.
In the $SU(5)$ case, the deviations from universality arise due to
the renormalization group running from the Planck scale
$M_{Pl}$ down to the $SU(5)$ scale $M_{GUT}$. Lepton flavor violation
diagrams arise due to the non-universality of the right-handed
slepton masses at the GUT scale. In contrast to the case we present here,
the main contributions to the $BR(\mu \rightarrow e \gamma)$ 
arise from the neutralino exchange diagram
amplitudes  $A_{M(n)}^L$, while  there is a similar 
cancellation to the one described in 
the above paragraph
due to the fact that the two main contributions enter
in the calculation with opposite sign. 
\footnote{In \cite{hisano}, the $LL$ and $LR$ diagrams
are discussed separately, while in ours
both contributions are included in the single neutralino exchange.}
In $SO(10)$ unified models \cite{Mario},
LFV arises due to the  non-universality of both right and left sleptons, and
hence this cancellation does not take place. 

We may further compare our results directly with the
results of \cite{KING}, where a similar effect is
observed, for the 
string-embedded version {ALR} of the
Pati-Salam model \cite{PS}
\footnote{The phenomenology of these string inspired models
has been discussed in \cite{STPH}.}.
In this work, large values for 
$\tan\beta$ are considered and  the contribution
arising from the chargino exchange diagram is the dominant one.
Moreover, in this 
analysis of $\tilde{\nu}_L$ terms,
Dirac masses in addition to the soft
left lepton masses are also 
included. For a certain value of $m_{3/2}$, the
renormalization-group  effects on the soft
masses are canceled by the Dirac masses, hence 
the masses of $\tilde{\nu}_L$  become
universal and the contribution  from the chargino diagram
vanishes. Then, the
$BR(\mu\rightarrow e +\gamma)$ decreases dramatically.
Although this effect agrees with our results, we 
would like to emphasize that the
small value of the $BR$ for certain values of $m_{3/2}$,
arises due to a cancellation
of two contributing  amplitudes. Note that the contribution of the Dirac
matrix to the sneutrino masses is absent in our analysis, as we have shown
in our treatment of the full neutral scalar mass.

Figs.\ref{fig:offm}, \ref{fig:offtb} and \ref{fig:offa}, show  the effects of
the changes of the input parameters in the total  
$BR(\mu\rightarrow e +\gamma)$. The changes in the non-diagonal elements 
of the scalar matrices can be induced from 
the approximate formulas that we presented,
however the behavior of the $BR$ is not correlated in all cases
to the increase in the mixing. More precisely:
\begin{itemize}
\item
Fig.\ref{fig:offm} shows the increase of the $BR$ as the 
gaugino masses decrease.
We can see that for a fixed value of $\tan\beta=7$,
and  values of $m_{1/2}$ leading to a SUSY mass
spectrum inside the experimental limits, the predicted values for the $BR$ 
are two  orders of magnitude lower than the experimental bounds.
\item
Fig.\ref{fig:offtb} shows the change of the predicted $BR$ for fixed values
of $m_{1/2}$. Here,
 we can see an increase of the $BR$ with $\tan\beta$ (solid
lines). The dashed lines are chosen
in a way that maximal $BR$'s are obtained;
still we can see that the $BR$'s stay
below the experimental bounds.
\item
Fig.\ref{fig:offa} shows the effect of $A_0$ in the
calculation. In this case, the 
behavior of the $BR$ is directly correlated to the increase in the scalar
mixings with  $A_0$. The sign of  $A_0$ has very little influence in the
total result, since it is the square of the
$A-$parameter that enters in the relevant renormalization-group
equations.
\end{itemize}

{\bf 2). Case with scalar mass  mixing at the GUT scale, $\Delta\neq 0$. }

Let us now proceed to discuss in detail
what happens in the case
that mixing of soft terms (arising through $\Delta_{L,R}\neq 0$) is 
already present at the GUT scale. Here we should note
that the contribution from 
$\Delta_{L,R}\neq 0$ (which is independent of
$\tan\beta$ and $m_{1/2}$), is much bigger than
$\delta m_N^2$. This implies that  there will be
a dramatic increase of all the Branching Ratios and all previously
noted effects have to be reconsidered.
Note for example that, since $m_{RR}^2$ is now non-diagonal,
additional amplitude contributions 
destroy the
cancellations that we observed in the previous
case.  Fig.~6 shows that  the absolute value of $A_M^L$ is 
now bigger than $A_M^R$,
(the contribution from  $A_M^L(c)$  to $A_M^L$ is again three orders of
magnitude smaller than the others,
 since it involves Yukawa couplings).

The modifications in the
case that mixing of soft-terms occurs at the
GUT scale, are presented in Figures \ref{fig:onm}, 
\ref{fig:ontb}. 
Changes of the $BR$ with $A_0$ are less relevant in this case since, 
as we mentioned before, the dominant mixing terms in the soft mass matrices are
independent of it. We have used  $A_0=-1.5 m_{3/2}$ in all the calculations 
of the rest of the section.

Figure \ref{fig:onm} indicates the variation of the
$BR$ as a function of $m_{3/2}$,
for $m_{1/2} = 200,300,500$ GeV and $\tan\beta =  7$.
As we see, for large values of 
$m_{3/2}$ the relevant branching ratio exhibits
a fixed-point behavior. The reason for this effect
is that, while $m_{1/2}$ enters in the calculation
via gaugino masses, $m_{3/2}$ multiplies the
scalar matrices and thus dominates the flavor-violating
processes in the case of non-zero off-diagonal
contributions at the GUT scale.

Figure \ref{fig:ontb} finally, shows the increase of the $BR$ as
$\tan\beta$ increases, for fixed 
values of  $m_{1/2} = 300$ GeV and $A = -1.5 m_{3/2}$ (solid lines).
Then, the experimental bounds impose severe constraints
on the allowed range of $(\tan\beta,m_{1/2},m_{3/2})$
that one may have. For certain combinations of
these terms, we predict $BR$ below the experimental limits. 
This means that the model we analyze, with non-universal soft terms
at the GUT scale, is consistent with the current experimental limits
for low masses of the sleptons and high masses for gauginos, when low
values of $\tan\beta$ are considered (dashed line).

Similar considerations may be done
for $\tau \rightarrow \mu \gamma$.
This is presented in fig.\ref{fig:tau}. Here,
the current experimental bounds are
not as strict as in the previous case.
Then, in the framework of the models that we are discussing,
we find that, for a wide region of the SUSY parameter
space, the predicted $BR$ is well below
the experimental bounds.
However, the prediction exceeds experimental limits
for larger values of $\tan\beta$.

Finally, in fig. \ref{fig:tracas} we 
compare our results with the choice of parameters
made in \cite{LT}. The choice of $b=1/2$ increases the values of the mixing
terms of the scalar matrices at the GUT scale as can be seen from
(\ref{eq:soft}), hence 
for this choice of 
parameters the results obtained for the $BR$ are one order of magnitude bigger
than in our case. 
 For completeness, we note that
in \cite{LT}, the choice of the 
Yukawa coefficients for the lepton matrix \ref{eq:lmass} were 
$c_{11}=4.0$, $c_{12}=c_{21}=0.9$, $c_{22}=1.08$, $c_{33}=1.9.$
In our case, we have instead 
$c_{11}=1.0$, $c_{12}=c_{21}=0.4$, $c_{22}=2.2$, $c_{33}=1.0$
and $b=1$. 

In retrospect, we can infer from the figures  
that in the case of non-universality in the scalar sector
the allowed ranges of $m_{1/2}$, $\tan\beta$ and $m_{3/2}$
are extremely limited. This is rather evident in particular from
our figure 8. Our conclusions are rather generic for this 
class of models, as far as mixing is also predicted 
in both the fermion and s-fermion mass textures.
 This fact naturally raises the question 
whether the simple $U(1)$-models are capable of generating
a completely realistic low energy  theory.   
We think that these problems cannot find a solution in the
present models. Again, as stressed also in the introduction,
flavor violations indicate that there should be a kind
of `mechanism' in the scalar sector to suppress large mixing
effects. In our opinion, string derived models are probably the only
realistic  ones which may  offer new `mechanisms' of additional
suppression. For example, in addition to the gauge and $U(1)$-family
symmetries, the trilinear and non-renormalizable terms in string
derived models have to respect additional symmetries arising
from modular invariance constraints. As a result, a large 
portion of ($U(1)$-invariant) mixing terms   are eliminated
by these string symmetries. Additional suppression of the 
mixing effects in the scalar sector may also arise by cyclic
symmetries as those discussed in ref\cite{AP}.
 
In the Figures, we have shown the 
constraints we can obtain in 
$m_{1/2}$, $\tan\beta$ and $m_{3/2}$
from $\mu \rightarrow e \gamma$ decays.
Let us finally  see what this implies for the
range of magnitudes of physical masses.
To do so, we give the low energy sparticle
masses for some indicative values of the input 
parameters: For a fixed value of 
$\tan\beta=7$, the lightest neutralino varies from 80 to
196 GeV as  $m_{1/2}$  
scales between  200 to 450 GeV.
For the same inputs, 
the lightest chargino varies from $196$ to $365$ GeV.
In both cases the dependence of the
results on  $m_{3/2}$ is small.
 For a fixed value 
$m_{1/2}=300$ GeV, the masses of the lightest neutralino and
chargino  change  by 
at most 3 GeV, while the masses of the
heavier charginos/neutralinos decrease
by up to 50 GeV, as 
$\tan\beta$ increases from 3 to 14.

What about the  charged slepton masses?
Here, we find that changes with $\tan\beta$
are  negligible for $m_{1/2} \approx 300$ GeV.
For $\tan\beta =7$ and $m_{1/2}$ = 200
GeV, when $m_{3/2}$ varies from 100 to 500 GeV, we
have:
$ m_{\tilde{\tau}_R} \approx (125-500)$ GeV,
$ m_{\tilde{e}_L} \approx (160-520)$ GeV,
while in this case the lightest scalar 
particle  for our initial conditions is
$ m_{\tilde{b}_R} \approx (100-500)$ GeV.
For $m_{1/2}$ = 450 GeV and the same range of 
$m_{3/2}$,
$ m_{\tilde{\tau}_R} \approx (200-520)$ GeV,
$ m_{\tilde{e}_L} \approx (300-590)$ GeV,
while the lightest scalar particle is again
$ m_{\tilde{b}_R} \approx (100-500)$ GeV.

\section{Conclusions}

In the present work, we investigated in detail
the predictions of $U(1)$ family symmetries 
for the rare processes $\mu\ra e\gamma$  and $\tau\ra 
\mu\gamma$.
We worked in the small $\tan\beta$ regime,
and found that in
 this class of models,
$\mu \ra e \gamma$ rare decays 
may occur at significant rates,
particularly in the case of non-zero
flavor mixing at the GUT scale.
Demanding that the predicted values do not exceed 
the current experimental limits,
we can put important bounds on the model
parameters. On the contrary,
$\tau\ra \mu\gamma$ decays 
are less dangerous and thus do not lead to
strong bounds.

Focusing on  $\mu \ra e \gamma$, 
we first discussed the case where no
mixing of soft terms at the GUT scale 
occurs. In this case lepton flavor violation
arises from non-zero neutrino masses 
in the theory. We find that
the dominant contributions
to the decay rate cancel around a certain value
of $m_{3/2}$, leading to very small values
for the relevant branching ratio. 
An important observation is that
the mixing effects of the Dirac matrix in
the light sneutrino sector are canceled
at first order.
Here, we should stress that once 
the accuracy of the experiments is improved,
it will be important to have a precise calculation which
takes into account 
the precise form of the sneutrino mass matrix.
The branching ratio of the process
increases  as $m_{1/2}$ becomes smaller
and  $\tan\beta$  larger, but 
we are still below the experimental bounds
for typical values of the model parameters.
Given that the experimental bounds will improve in the future,
non-detection of
$\mu\rightarrow e \gamma$ events will be 
associated to low values of
gaugino masses as we can see in figure~\ref{fig:offtb}.

The situation changes when 
mixing effects in the soft masses are introduced
at the GUT scale. Since the effects of the
off-diagonal terms are much larger than
those of the massive neutrinos, the
cancellation effects that we discussed
are no longer present and
large lepton number violating effects 
may be generated. 
As before, the branching ratio of the decay
increases  as $m_{1/2}$ becomes smaller
and  $\tan\beta$  larger, but now strong bounds
on the model parameters are derived.
In this scenario, the $BR(\mu\rightarrow e \gamma)$ experimental bounds
constrain the SUSY parameter
space to large values for gaugino masses and small scalar
masses as shown in figure~\ref{fig:ontb}.

{\bf{Acknowledgments}}
This work is partially supported by the TMR contract 
``Beyond the Standard Model'', 
ERBFMRX-CT96-0090.
MEG wishes to acknowledge  useful discussions with A. Dedes regarding
the numerical integration of the RG equations.
MEG and GKL  wish to thank the CERN theory division 
where part of this work has been done,
for kind hospitality.


\newpage
\begin{figure}
\begin{center}
\epsfig{file=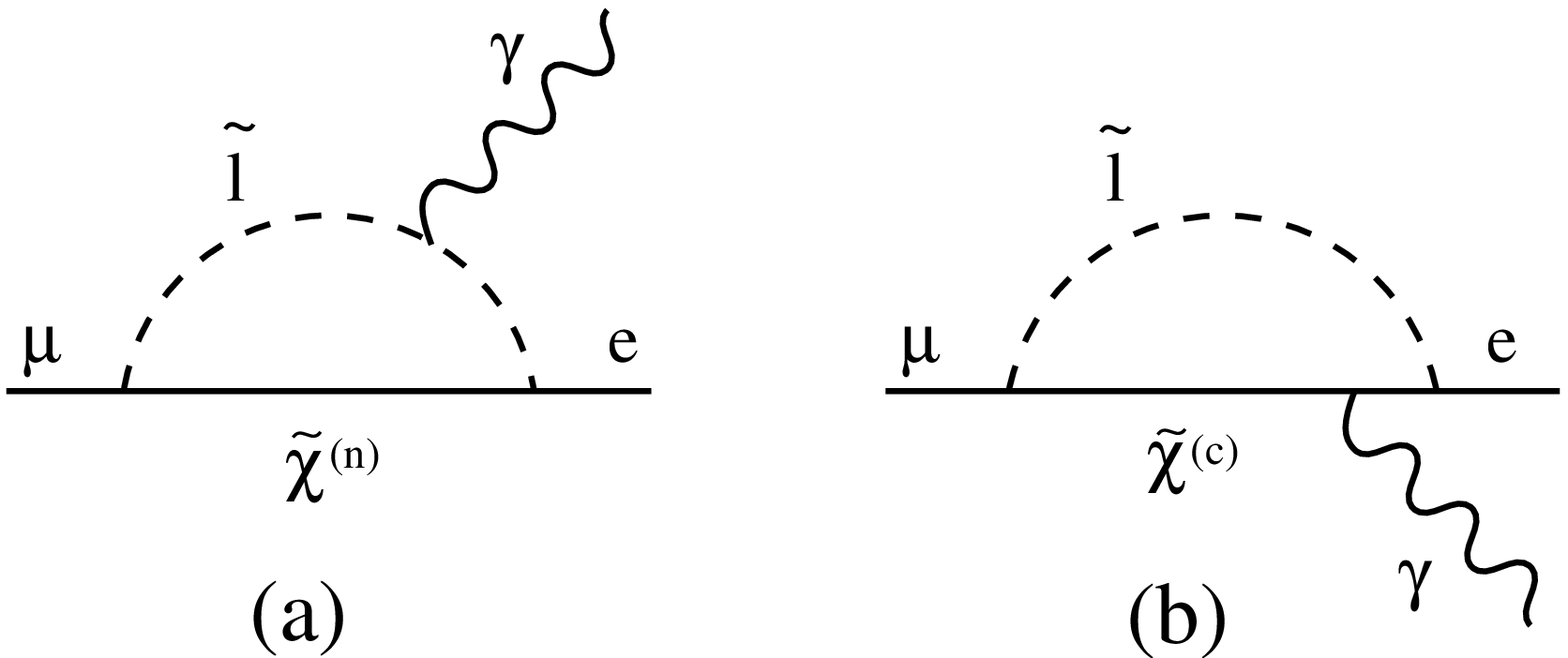,width=14cm}
\end{center}
\caption{The generic Feynman diagrams for the $\mu\ra e\gamma$
decay. $\tilde l$ stands for charged slepton (a) or sneutrino (b), while
$\tilde\chi ^{(n)}$ and $\tilde\chi ^{(c)}$ represent  
neutralinos and charginos respectively.}
\label{figure1}
\end{figure}

\begin{figure}

\begin{center}
\hspace{-1cm}
\epsfig{figure=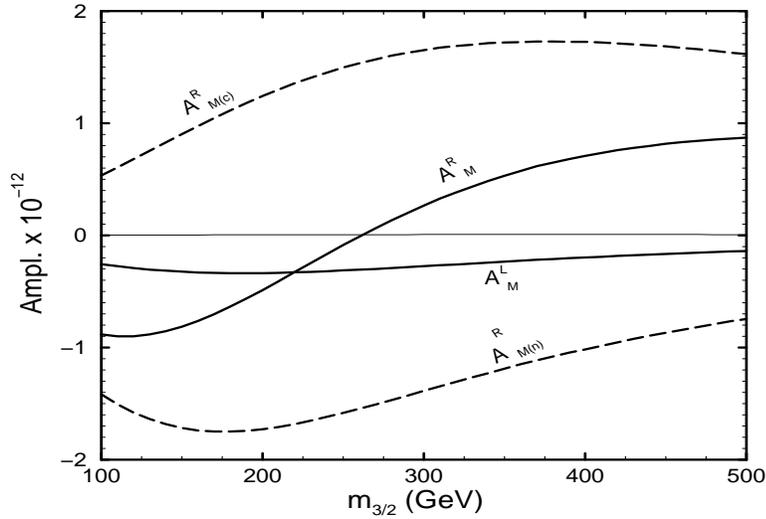,height=7truecm,width=10truecm}
\vspace{.5cm}
\caption{Amplitudes contributing to the $BR$$(\mu\rightarrow e \gamma)$ when 
universal soft masses at the GUT scale are considered ($\Delta =0$). Dashed 
lines are the two contributions to $A_M^R$. The curves 
are obtained using
$\tan\beta=7$, $m_{1/2}=300$ GeV, and $A_0=-1.5 m_{3/2}$ as input parameters.}
\end{center}
\label{fig:ampoff1}
\end{figure}

\begin{figure}

\begin{center}
\hspace*{-1cm}
\epsfig{figure=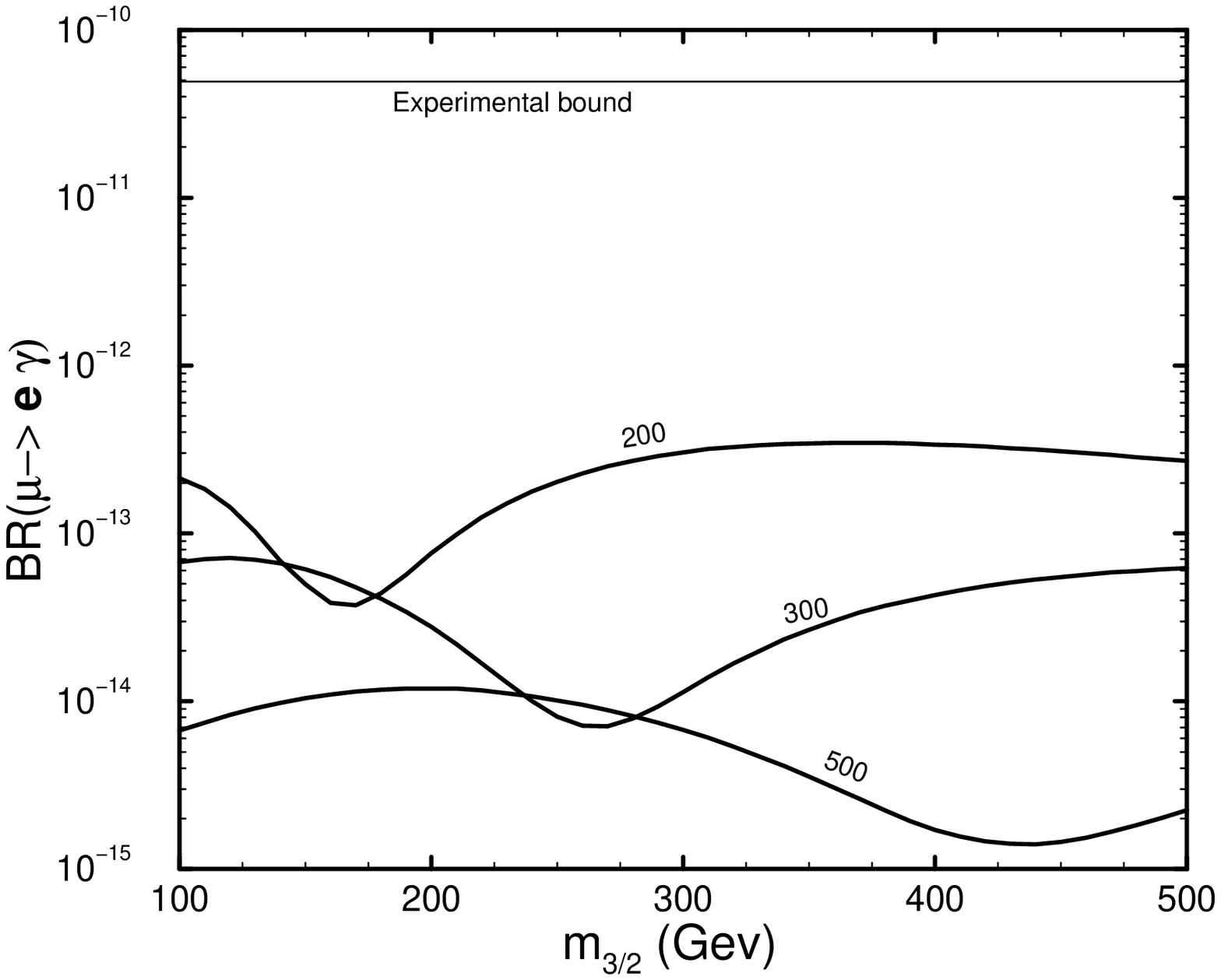,height=7truecm,width=10truecm}
\vspace{.5cm}
\caption{$BR$$(\mu\rightarrow e \gamma)$ for a range of values of
$m_{1/2}$ (labeled above). Universal soft masses at the GUT scale 
are considered ($\Delta = 0$). 
The curves are obtained using 
$\tan\beta=7$ and $A_0=-1.5 m_{3/2}$ as input parameters.}
\label{fig:offm}
\end{center}

\end{figure}


\begin{figure}

\begin{center}
\hspace*{-1cm}
\epsfig{figure=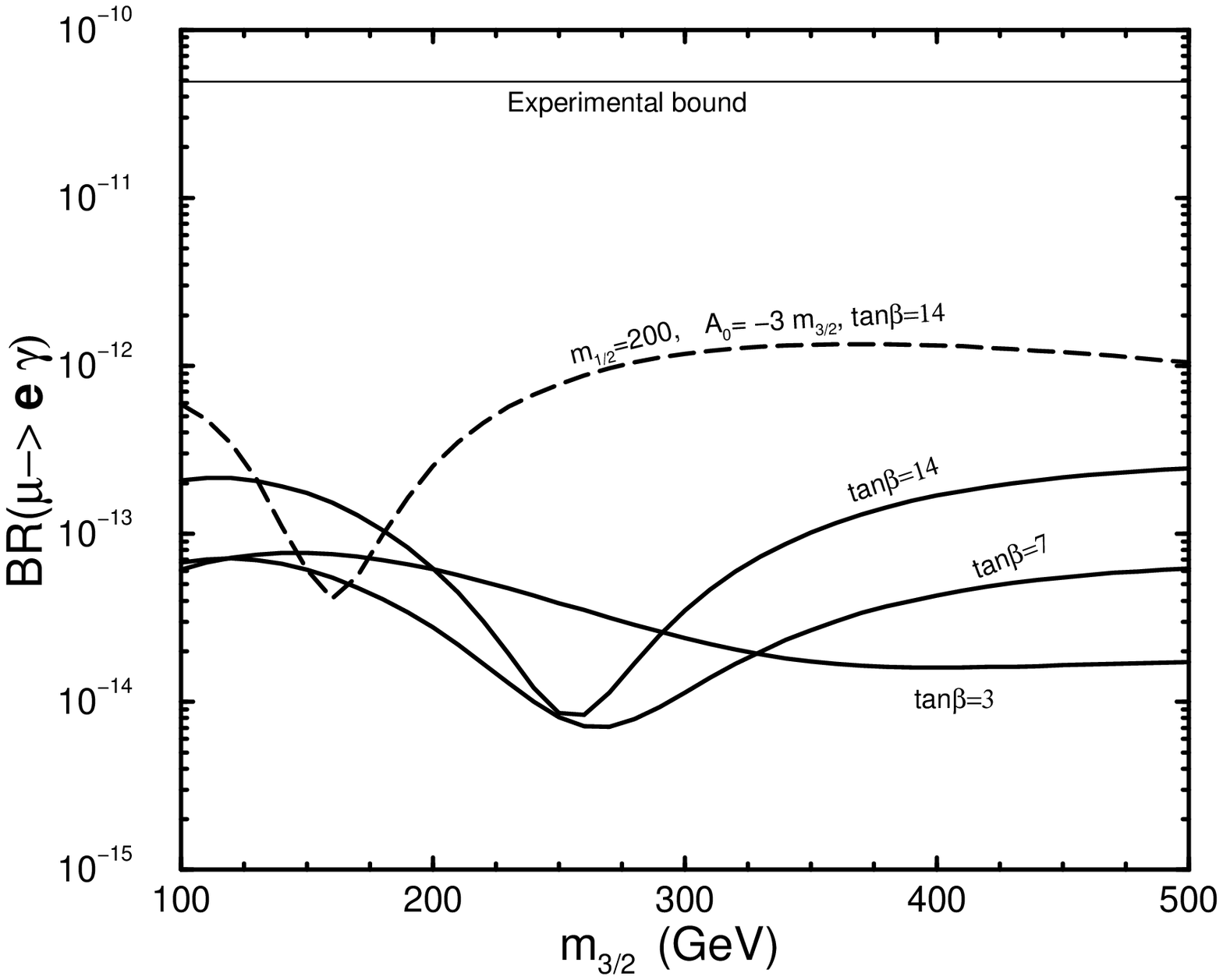,height=7truecm,width=10truecm}
\vspace{.5cm}
\caption{$BR$$(\mu\rightarrow e \gamma)$ for a range of values of
$\tan\beta$ (labeled above). Universal soft masses at the GUT scale 
are considered ($\Delta = 0$). 
Solid lines are obtained using 
 $m_{1/2}=300$ GeV and $A_0=-1.5 m_{3/2}$ as input parameters.}
\label{fig:offtb}
\end{center}

\end{figure}


\begin{figure}

\begin{center}
\hspace*{-1cm}
\epsfig{figure=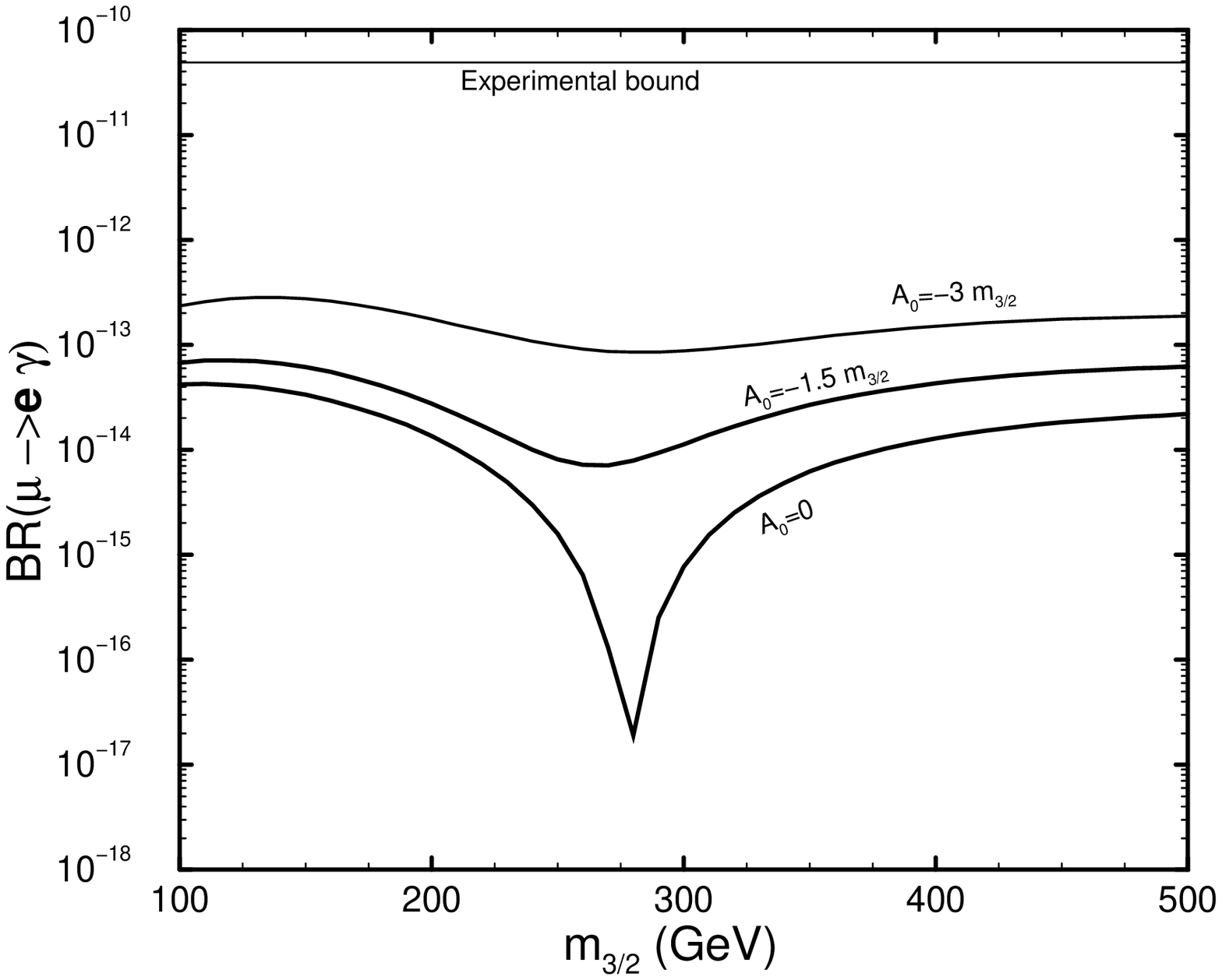,height=7truecm,width=10truecm}
\vspace{.5cm}
\caption{$BR$$(\mu\rightarrow e \gamma)$ for a range of values of
$A_0$ (labeled above). Universal soft masses at the GUT scale 
are considered ($\Delta= 0$). 
The curves are obtained using 
$\tan\beta=7$ and $m_{1/2}=300$ GeV as input parameters.
}
\label{fig:offa}
\end{center}

\end{figure}
\begin{figure}

\begin{center}
\hspace*{-1cm}
\epsfig{figure=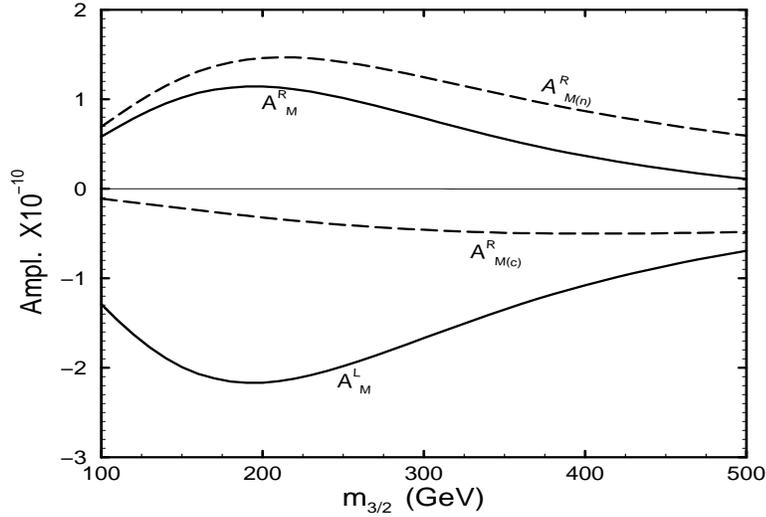,height=7truecm,width=10truecm}
\vspace{.5cm}
\caption{Amplitudes contributing to the $BR$$(\mu\rightarrow e \gamma)$ when 
non universal soft masses at the GUT scale are considered ($\Delta \neq 0$). 
The curves are obtained using 
$\tan\beta=7$, $m_{1/2}=300$ GeV, and $A_0=-1.5 m_{3/2}$ as input parameters.}
\end{center}
\label{fig:amon}
\end{figure}

\begin{figure}

\begin{center}
\hspace*{-1cm}
\epsfig{figure=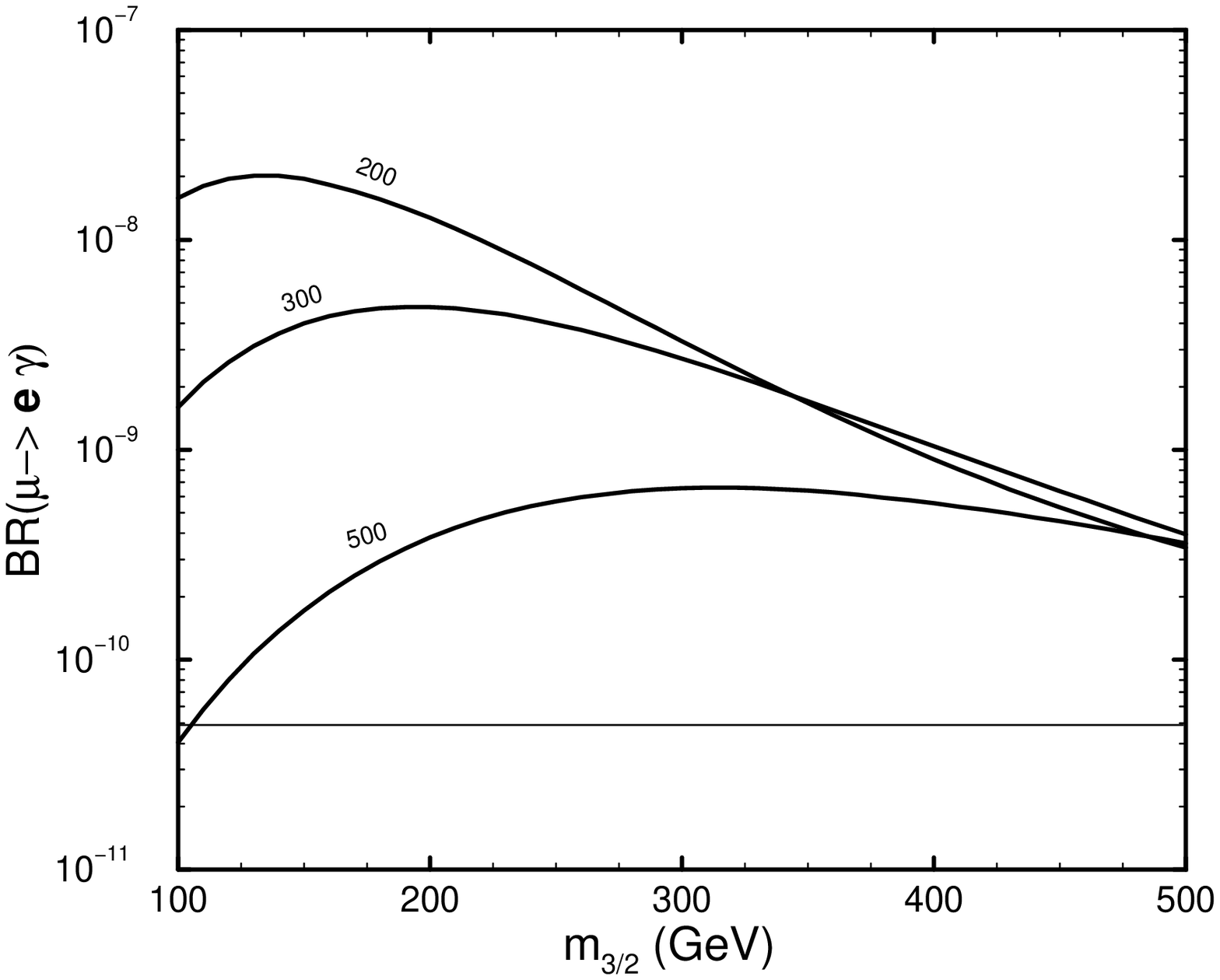,height=7truecm,width=10truecm}
\vspace{.5cm}
\caption{$BR$$(\mu\rightarrow e \gamma)$ for a range of values of
$m_{1/2}$ (labeled above). Non universal  
soft masses at the GUT scale 
are considered ($\Delta \neq 0$). 
The curves are obtained using 
$\tan\beta=7$ and $A_0=-1.5 m_{3/2}$ as input parameters.}
\label{fig:onm}
\end{center}

\end{figure}


\begin{figure}

\begin{center}
\hspace*{-1cm}
\epsfig{figure=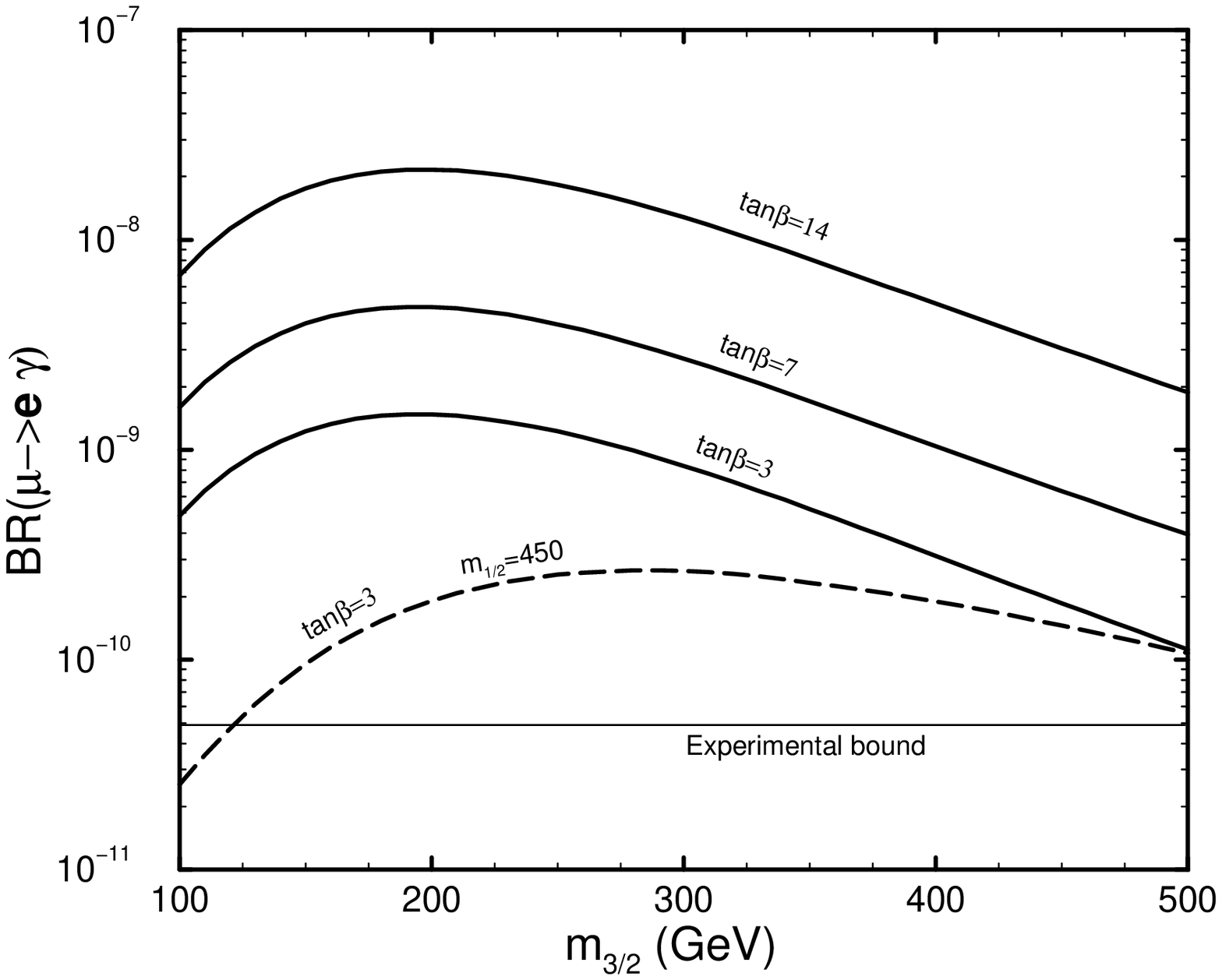,height=7truecm,width=10truecm}
\vspace{.5cm}
\caption{$BR$$(\mu\rightarrow e \gamma)$ for a range of values of
$\tan\beta$ (labeled above). Non universal  
soft masses at the GUT scale 
are considered ($\Delta \neq 0$). 
Solid lines are obtained using, $m_{1/2}=300$ GeV, and $A_0=-1.5 m_{3/2}$ as 
input parameters.}
\label{fig:ontb}
\end{center}

\end{figure}


\begin{figure}
\begin{center}
\hspace*{-1cm}
\epsfig{figure=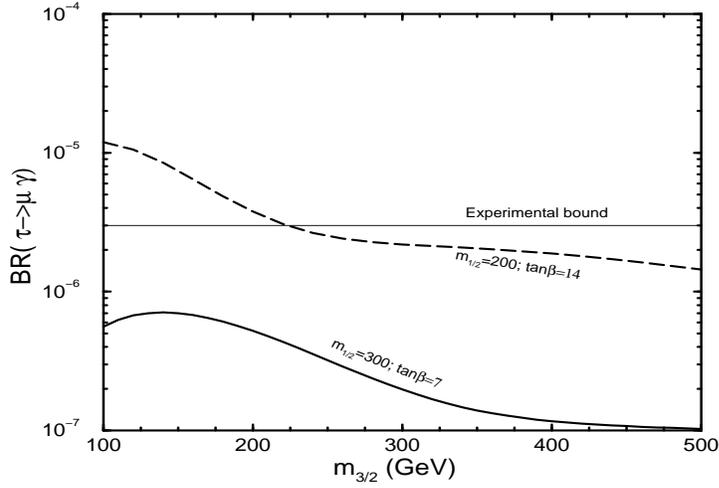,height=6.5truecm,width=9.5truecm}
\vspace{.5cm}
\caption{$BR$$(\tau\rightarrow \mu \gamma)$. Non universal  
soft masses at the GUT scale 
are considered ($\Delta \neq 0$). Experimental limits are violated for large
$\tan\beta$ and small values of $m_{3/2}$ (dashed line). In both cases
 $A_0=-1.5 m_{3/2}$.}
\label{fig:tau}
\end{center}
\end{figure}

\begin{figure}
\begin{center}
\hspace*{-1cm}
\epsfig{figure=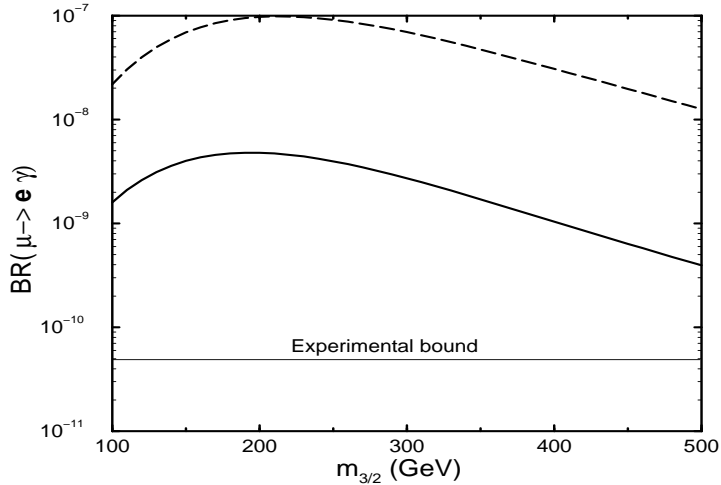,height=6.5truecm,width=9.5truecm}
\vspace{.5cm}
\caption{$BR$$(\tau\rightarrow \mu \gamma)$, for $m_{1/2}=300$ GeV, 
$\tan\beta=7$,
and $A_0=-1.5 m_{3/2}$. The dashed line  is  obtained 
using the charge and coefficient choices of \cite{LT},
while the solid line  is  derived using
the choice of parameters presented in this paper.
 Non universal  
soft masses at the GUT scale 
are considered ($\Delta \neq 0$)                                                                                .}
\label{fig:tracas}
\end{center}
\end{figure}

\end{document}